# Estimating the COVID-19 Infection Rate: Anatomy of an Inference Problem

Charles F. Manski
Department of Economics and Institute for Policy Research, Northwestern University

and

Francesca Molinari
Department of Economics, Cornell University

draft: April 13, 2020

Abstract

As a consequence of missing data on tests for infection and imperfect accuracy of tests, reported rates of population infection by the SARS CoV-2 virus are lower than actual rates of infection. Hence, reported rates of severe illness conditional on infection are higher than actual rates. Understanding the time path of the COVID-19 pandemic has been hampered by the absence of bounds on infection rates that are credible and informative. This paper explains the logical problem of bounding these rates and reports illustrative findings, using data from Illinois, New York, and Italy. We combine the data with assumptions on the infection rate in the untested population and on the accuracy of the tests that appear credible in the current context. We find that the infection rate might be substantially higher than reported. We also find that the infection fatality rate in Italy is substantially lower than reported.

Acknowledgements: We thank Yizhou Kuang for able research assistance. We thank Michael Gmeiner, Valentyn Litvin, and Jörg Stoye for helpful comments. We are grateful for the opportunity to present this work at an April 13, 2020 virtual seminar at the Institute for Policy Research, Northwestern University.

*Assumptions + Data  ⇨ Conclusions*

1. Introduction

It is well appreciated that accurate characterization of the time path of the coronavirus pandemic has been hampered by a serious problem of missing data. Confirmed cases have commonly been measured by rates of positive findings among persons who have been tested for infection. Infection data are missing for persons who have not been tested. It is also well-appreciated that the persons who have been tested differ considerably from those who have not been tested. Criteria used to determine who is eligible for testing typically require demonstration of symptoms associated with presence of infection or close contact with infected persons. This gives considerable reason to believe that some fraction of untested persons are asymptomatic or pre-symptomatic carriers of the COVID-19 disease. Presuming this is correct, the actual rate of infection has been higher than the reported rate.

It is perhaps less appreciated that available measurement of confirmed cases is imperfect because the prevalent tests for infection are not fully accurate. There is basis to think that accuracy is highly asymmetric. Various sources suggest that the positive predictive value (the probability that, conditional on testing positive, an individual is indeed infected) of the tests in use is close to one.  However, it appears that the negative predictive rate (the probability that, conditional on testing negative, the individual is indeed not infected) may be substantially less than one. Presuming this asymmetry, the actual rate of infection has again been higher than the reported rate.

Combining the problems of missing data and imperfect test accuracy yields the conclusion that reported rates of infections are lower than actual rates. Reported rates of infection have been used as the denominator for computation of rates of severe disease conditional on infection, measured by rates of hospitalization, treatment in intensive care units (ICUs), and death. Presuming that the numerators in rates of severe illness conditional on infection have been measured accurately, reported rates of severe illness conditional on infection are higher than actual rates.

On March 3, 2020 the Director General of the World Health Organization (WHO) stated:[1] "Globally, about 3.4% of reported COVID-19 cases have died." It is tempting to interpret the 3.4% number as the actual case-fatality ratio (CFR). However, if deaths have been recorded accurately and if the actual rate of infection has been higher than the reported rate, the WHO statistic should be interpreted as an upper bound on the actual CFR on that date. Recognizing this, researchers have recommended random testing of populations as a potential future method to solve the missing data problem.[2]

In the present absence of random testing, various researchers have put forward point estimates and forecasts for infection rates and rates of severe illness derived in various ways. Work performed by separate groups of epidemiologists at the Imperial College COVID-19 Response Team and the University of Washington's Institute for Health Metrics and Evaluation has received considerable public attention.[3] The available estimates and forecasts differ in the assumptions that they use to yield specific values. The assumptions vary substantially and so do the reported findings. To date, no particular assumption or resulting estimate has been thought sufficiently credible as to achieve consensus across researchers.

We think it misguided to report point estimates obtained under assumptions that are not well justified. We think it more informative to determine the range of infection rates and rates of severe illness implied by a credible spectrum of assumptions. In some disciplines, research of this type is called sensitivity analysis. A common practice has been to obtain point estimates under alternative strong assumptions. A problem with sensitivity analysis as usually practiced is that, in many applications, none of the strong assumptions entertained has a good claim to realism.

Rather than perform traditional sensitivity analysis, this paper brings to bear econometric research on partial identification. Study of partial identification analysis removes the focus on point estimation obtained under strong assumptions. Instead it begins by posing relatively weak assumptions that should be highly credible in the applied context under consideration. Such weak assumptions generally imply set-valued estimates rather than point estimates. Strengthening the initial weak assumptions shrinks the size of the implied set estimate. The formal methodological problem is to determine the set estimate that logically results when available data are combined with specified assumptions. See Manski (1995, 2003, 2007) for monograph expositions at different technical levels. See Tamer (2010) and Molinari (2020) for review articles.

---

[1] https://www.who.int/dg/speeches/detail/who-director-general-s-opening-remarks-at-the-media-briefing-on-covid-19---3-march-2020
[2] See, for example, https://www.statnews.com/2020/03/17/a-fiasco-in-the-making-as-the-coronavirus-pandemic-takes-hold-we-are-making-decisions-without-reliable-data/
[3] See https://www.imperial.ac.uk/mrc-global-infectious-disease-analysis/covid-19/ and http://www.healthdata.org/ respectively.

Considering estimation of the infection rate for the coronavirus, combining available data with credible assumptions easily yields a lower bound on the infection rate. The harder problem is to determine an upper bound that is both credible and informative, so as to obtain an interval estimate that is credible and informative. This paper explains the logic of the identification problem, determines the identifying power of some credible assumptions, and reports illustrative set estimates.

We analyze data from Illinois, New York, and Italy, over the period March 16 to April 6, 2020. We impose weak monotonicity assumptions on the rate of infection in the untested sub-population to draw credible conclusions about the population infection rate. We find that the infection rate as of April 6, 2020, for Illinois, New York, and Italy are, respectively, bounded in the intervals [0.001, 0.517], [0.008, 0.645], and [0.003, 0.510]. Further analyzing the rate of hospitalization, treatment in an intensive care unit, and death in Italy, we find that as of April 6 the rates for these severe outcomes are bounded, respectively, in the intervals [0.001, 0.172], [0, 0.02], and [0.001, 0.086]. The upper bound on the fatality rate is substantially lower than that among confirmed infected individuals, which was 0.125 on April 6.

2. Methods

We first address the basic problem of bounding the infection rate as of a specified date. The analysis initially considers the problem abstractly and then derives bounds under particular credible assumptions. We next show how bounding the infection rate yields a bound on the rate of severe illness conditional on infection. We then extend the analysis to bound the rates conditional on observed patient characteristics. Knowledge of all of these rates is important to inform public health policy.

2.1. Bounding the Population Infection Rate

Consider a specified population of persons who are alive at the onset of the pandemic. This may, for example, be the population of a city, state, or nation. Let the objective be to determine the fraction of the population who are infected by the SARS-CoV-2 virus by a specified date d. Synonymously, this is the fraction of the population who experience onset of the COVID-19 disease by date d.

The present analysis assumes that a person can have at most one disease episode. If a person has been infected, the person either achieves immunity after recovery or dies. The assumption that immunity is achieved after recovery, at least for some period of time, is consistent with current knowledge of the disease. We also assume that the size of the population is stable over the time period of interest. Thus, we abstract from the fact that, as time passes, deaths from the disease and other causes reduce the size of the population, births increase the size, and migration may reduce or increase the size on net.

Let $C_d = 1$ if a person has been infected by the coronavirus by date d and $C_d = 0$ otherwise. The objective is to determine $P(C_d = 1)$, the probability that a member of the population has been infected by date d. Equivalently, $P(C_d = 1)$ is the infection rate or the fraction of persons who have experienced disease onset. Our analysis of the problem of inference on $P(C_d = 1)$ is a simple extension of ideas that have been used regularly in the literature on partial identification, beginning with study of inference with missing outcome data (Manski, 1989).

$P(C_d = 1)$ is not directly observable. However, population surveillance systems provide daily data on two quantities related to $P(C_d = 1)$. These are the rate of testing for infection and the rate of positive results among those tested. To simplify analysis, we assume that a person is tested at most once by date d. This assumption may not be completely accurate, for reasons that will be explained later.

Let $T_d = 1$ if a person has been tested by date d and $T_d = 0$ otherwise. Let $R_d = 1$ if a person has received a positive test result by date d and $R_d = 0$ otherwise. Observe that $T_d = 0 \Rightarrow R_d = 0$ and $R_d = 1 \Rightarrow T_d = 1$. By the Law of Total Probability, the infection rate may be written as follows:

(1) $\quad P(C_d = 1) \ = \ P(C_d = 1|R_d = 1)P(R_d = 1) + P(C_d = 1| R_d = 0)P(R_d = 0),$

where

(2) $\quad P(R_d = 1) \ = \ P(R_d = 1|T_d = 1)P(T_d = 1),$

(3) $\quad P(R_d = 0) \ = \ P(T_d = 0) \ + P(R_d = 0|T_d = 1)P(T_d = 1),$

(4) $\quad P(C_d = 1| R_d = 0)P(R_d = 0) \ = \ P(C_d = 1, R_d = 0)$
$\quad\quad\quad\quad\quad = \ P(C_d = 1|T_d = 0)P(T_d = 0) \ + P(C_d = 1|T_d = 1, R_d = 0)P(R_d = 0|T_d = 1)P(T_d = 1).$

Now consider each of the component quantities that together determine the infection rate. Assuming that reporting of testing is accurate, daily surveillance reveals the testing rate and the rate of positive results among those tested. Thus, the quantities $P(T_d = 0)$, $P(T_d = 1)$, $P(R_d = 0|T_d = 1)$, and $P(R_d = 1|T_d = 1)$ are directly observable. The remaining quantities are not directly observable.

The quantities $P(C_d = 1|R_d = 1)$ and $P(C_d = 1|T_d = 1, R_d = 0)$ are determined by the accuracy of testing. The former is the positive predictive value (PPV) and the latter is one minus the negative predictive value (NPV). We note that medical researchers and clinicians often measure test accuracy in a different way, through test sensitivity and specificity. The sensitivity and specificity of tests for COVID-19 on the tested sub-population are $P(R_d = 1|T_d = 1, C_d = 1)$ and $P(R_d = 0|T_d = 1, C_d = 0)$ respectively. Sensitivity and specificity are related to PPV and NPV through Bayes Theorem, whose application generally requires

knowledge of $P(C_d = 1 | T_d = 1)$, the infection rate in the tested sub-population. An exception to this generalization is that PPV equals one if and only if specificity equals one, whenever $P(C_d = 1 | T_d = 1) > 0$.

Medical experts believe that the PPV of the prevalent tests for COVID-19 is close to one, but that NPV may be considerably less than one. We have obtained this information in part from personal communication with an infectious disease specialist at Northwestern Memorial Hospital and in part from the public literature. For example, USA Today has reported as follows:[4]

> "Dwayne Breining, executive director for Northwell Labs in New Hyde Park, New York, said the test is extremely accurate and can detect even low levels of the virus. False positives are highly unlikely, he said, though false negatives may result from poor-quality swabs or if the instrument is blocked by mucus. Those factors might have been at play in a number of false negatives initially reported. Patients who continue to have symptoms after a negative test are advised to get retested."

We therefore find it credible to assume that $P(C_d = 1 | R_d = 1) = 1$. It can be shown that this is equivalent to assuming that test specificity $P(R_d = 0 | T_d = 1, C_d = 0) = 1$. The final sentence of the Breining quote explains why it may not be completely accurate to assume that persons are tested at most once, but we maintain this assumption for simplicity.

There does not appear to presently be a firm basis to determine the precise NPV of the prevalent nasal-swab tests, but there may be a basis to determine a credible bound. Medical experts have been cited as believing that the rate of false-negative test findings is at least 0.3. However, it is not clear whether they have in mind one minus the NPV or one minus test sensitivity.[5] One may perhaps find it credible to extrapolate from experience testing for influenza to testing for covid-19. For example, Peci *et al*. (2014) study the performance of rapid influenza diagnostic testing. They find a PPV of 0.995 and an NPV of 0.853.

It is not clear whether NPV has been constant over the short time period we study or, contrariwise, has varied as testing methods and the subpopulation of tested persons change over time. The NPV may also vary over longer periods if the virus mutates significantly. The illustrative results that we report later assume that NPV is in the range [0.6, 0.9], implying that $P(C_d = 1 | T_d = 1, R_d = 0) \in [0.1, 0.4]$.[6]

---

[4] https://www.usatoday.com/story/news/nation/2020/03/16/coronavirus-what-expect-when-you-get-tested-covid-19/5061120002/
[5] See, for example, https://www.livescience.com/covid19-coronavirus-tests-false-negatives.html.

It remains to consider $P(C_d = 1 | T_d = 0)$, the rate of infection among those who have not been tested. This quantity has been the subject of much discussion, with substantial uncertainty expressed about its value. It may be that the value changes over time as criteria for testing people evolve and testing becomes more common. The illustrative results that we report later show numerically how the conclusions one can draw about $P(C_d = 1)$ depends on the available knowledge of $P(C_d = 1 | T_d = 0)$.

To finalize the logical derivation of a bound on $P(C_d = 1)$, let $[L_{d0}, U_{d0}]$ and $[L_{d10}, U_{d10}]$ denote credible lower and upper bounds on $P(C_d = 1 | T_d = 0)$ and $P(C_d = 1 | T_d = 1, R_d = 0)$ respectively. Now combine these bounds with the assumption that $P(C_d = 1 | R_d = 1) = 1$ and with empirical knowledge of the testing rate and the rate of positive test results. Then equations (1) – (4) imply this bound on the population infection rate:

(5) $P(R_d = 1) + L_{d0}P(T_d = 0) + L_{d10}P(R_d = 0 | T_d = 1)P(T_d = 1) \leq P(C_d = 1)$

$$\leq P(R_d = 1) + U_{d0}P(T_d = 0) + U_{d10}P(R_d = 0 | T_d = 1)P(T_d = 1).$$

The width of bound (5) is

(6) $(U_{d10} - L_{d10})P(R_d = 0 | T_d = 1)P(T_d = 1) + (U_{d0} - L_{d0})P(T_d = 0)$.

Inspection of (6) shows that uncertainty about test accuracy and about the infection rate in the untested sub-population, measured by $U_{d10} - L_{d10}$ and $U_{d0} - L_{d0}$, combine linearly to yield uncertainty about the population infection rate. The fractions $P(T_d = 1)$ and $P(T_d = 0)$ of the population who have and have not been tested linearly determine the relative contributions of the two sources of uncertainty.

---

[6] Rather than assume a bound on NPV directly, a medical researcher or clinician could assume a bound on sensitivity. It can be shown that a bound on sensitivity combined with the assumption that specificity equals one implies a bound on the NPV.

## 2.2. Using Monotonicity Assumptions to Bound the Infection Rate of Untested Persons

As of early April 2020, the fraction of the population who have been tested is very small in most locations. For example, the fraction who have been tested by April 6, 2020 was about 0.005 in Illinois, 0.017 in New York, and 0.012 in Italy; see Section 3 for details on the data sources. Hence, the present dominant concern is uncertainty about the infection rate in untested sub-populations. We now consider the problem of obtaining a credible bound on this quantity.

The worst case from the perspective of uncertainty would occur if society possesses no credible information about the value of $P(C_d = 1 | T_d = 0)$, so $U_{d0} - L_{d0} = 1$. Then the bound on the population infection rate has width no smaller than $P(T_d = 0)$, even if test accuracy is known precisely. In the three locations and on the date mentioned above, the widths of the bounds on the population infection rate are greater than (0.995, 0.983, 0.988). Thus, the available data on the rate of positive tests for tested persons reveal almost nothing about the population infection rate. Moreover, a huge increase in the rate of testing would be required to substantially narrow the width of the bound.

The best case would occur if society somehow were to possess precise credible knowledge of $P(C_d = 1 | T_d = 0)$. Then $U_{d0} - L_{d0} = 0$ and the bound on the population infection rate has width $(U_{d10} - L_{d10}) P(R_d = 0 | T_d = 1) P(T_d = 1)$. In the current setting, where $P(T_d = 1)$ is very small, this width has negligible magnitude even if a large fraction of tested persons have negative results and there is great uncertainty about test accuracy.

We judge the current situation to be intermediate between the worst and best case scenarios. We are aware of no credible way to assign a precise value to $P(C_d = 1 | T_d = 0)$, nor even to place a tight bound on the quantity. On the other hand, it is too pessimistic to view society as having no relevant information. Two monotonicity assumptions are highly credible in the current context.

*Testing Monotonicity*

Present criteria for testing persons for infection by the coronavirus commonly require the person to display symptoms of infection or to have been in close contact with someone who has tested positive. These criteria strongly suggest that, as of each date d, the infection rate among tested persons is higher than the rate among untested persons. This yields the *testing-monotonicity* assumption

(7) $\quad P(C_d = 1|T_d = 0) \leq P(C_d = 1|T_d = 1)$.

Observe that if testing for infection were random rather than determined by the current criteria, it would be credible to impose a much stronger assumption, namely $P(C_d = 1|T_d = 0) = P(C_d = 1|T_d = 1)$. However, testing clearly has not been random. Hence, we only impose assumption (7).

Research on partial identification has often exploited monotonicity assumptions similar to (7), beginning with Manski and Pepper (2000). To use the assumption in the present setting, consider the quantity $P(C_d = 1|T_d = 1)$. The Law of Total Probability, the maintained assumption that positive test results are always accurate, and the specified upper bound on $P(C_d = 1|T_d = 1, R_d = 0)$ yield

$$
\begin{aligned}
(8) \quad P(C_d = 1|T_d = 1) &= P(R_d = 1|T_d = 1) + P(C_d = 1|T_d = 1, R_d = 0)P(R_d = 0|T_d = 1) \\
&\leq P(R_d = 1|T_d = 1) + U_{d10}P(R_d = 0|T_d = 1).
\end{aligned}
$$

Combining (7) and (8) yields this upper bound on $P(C_d = 1|T_d = 0)$.

(9) $\quad U_{d0} = P(R_d = 1|T_d = 1) + U_{d10}P(R_d = 0|T_d = 1) = U_{d10} + (1 - U_{d10})P(R_d = 1|T_d = 1)$.

Bound (9) is methodologically interesting because $U_{d0}$ is now a function of $U_{d10}$ rather than a separate quantity. It thus enhances the importance of securing an informative upper bound on $P(C_d = 1|T_d = 1, R_d =$ 0). In particular, (9) implies that $U_{d0} \geq U_{d10}$, whatever the rate $P(R_d = 1|T_d = 1)$ of positive test outcomes may be.

The monotonicity assumption does not affect the lower bound $L_{d0}$, which is zero in the absence of other information. Hence, inserting $L_{d0} = 0$ and (9) into the bound (5) on $P(C_d = 1)$ yields

(10) $P(R_d = 1) + L_{d10}P(R_d = 0|T_d = 1)P(T_d = 1) \leq P(C_d = 1)$

$\leq P(R_d = 1) + U_{d10}P(R_d = 0|T_d = 1)P(T_d = 1) + [P(R_d = 1|T_d = 1) + U_{d10}P(R_d = 0|T_d = 1)]P(T_d = 0)$.

The width of bound (10) is

(11) $(U_{d10} - L_{d10})P(R_d = 0|T_d = 1)P(T_d = 1) + [P(R_d = 1|T_d = 1) + U_{d10}P(R_d = 0|T_d = 1)]P(T_d = 0)$.

In the present context where $P(T_d = 1)$ is very small, the width of the bound approximately equals the sum of the rate $P(R_d = 1|T_d = 1)$ of positive test results plus the product of the rate $P(R_d = 0|T_d = 1)$ of negative test results and the upper bound on $P(C_d = 1|T_d = 1, R_d = 0)$.

*Temporal Monotonicity*

A second form of monotonicity holds logically rather than by assumption. Our analysis thus far has only considered the infection rate by a specified date. A person who has been infected by an early date necessarily has been infected by every later date. Hence, for two dates d and d', we have the *temporal monotonicity* condition

(12) $d' < d \Rightarrow P(C_{d'} = 1) \leq P(C_d = 1)$.

Inequality (12) makes date a monotone *instrumental variable* as defined in Manski and Pepper (2000). Proposition 1 of that article shows that, given a set of date-specific lower and upper bounds on the infection rate for various dates, condition (12) implies that $P(C_d = 1)$ must be greater than or equal to the maximum of the date-specific lower bounds for all d' ≤ d. Moreover, $P(C_d = 1)$ must be less than or equal to the minimum of the date-specific upper bounds for all d' ≥ d.[7] Applying this result to the date-specific bounds (10) yields this result:

$$(13) \quad \max_{d' \leq d} P(R_{d'} = 1) + L_{d'10} P(R_{d'} = 0 | T_{d'} = 1) P(T_{d'} = 1) \leq P(C_d = 1)$$

$$\leq \min_{d' \geq d} P(R_{d'} = 1) + U_{d'10} P(R_{d'} = 0 | T_{d'} = 1) P(T_{d'} = 1)$$

$$+ [P(R_{d'} = 1 | T_{d'} = 1) + U_{d'10} P(R_{d'} = 0 | T_{d'} = 1)] P(T_{d'} = 0).$$

Bound (13) necessarily is a subset of bound (10). The lower bound in (13) equals the lower bound in (10) if the lower bound in (10) at all dates d' < d is always less than or equal to the lower bound at d. If this does not hold, as can happen with some configurations of testing data, the lower bound on (13) is greater than the one in (10). Symmetrically, the upper bound in (13) equals the upper bound (10) if the upper bound in (10) at all dates d' > d is always greater than or equal to the upper bound at d. If this does not hold, the upper bound on (13) is less than the one in (10). Thus, the temporal monotonicity condition may or may not have identifying power, depending on the testing data. We find that it modestly improves lower bounds with the data we use.

---

[7] Proposition 1 of Manski and Pepper (2000) shows that this bound is sharp. That is, it is the tightest bound achievable with the available information. Molinari (2020, Section 2.1) shows that it is a more complex matter to obtain sharp bounds for functions of the infection rate that vary with time.

2.3. Bounding the Fraction of Asymptomatic Infections

We are presently unaware of other assumptions or logical conditions that enjoy credibility comparable to the above monotonicity assumptions and that have identifying power. One may, however, perhaps feel comfortable bringing to bear assumptions whose credibility stems from the judgement of respected medical and epidemiological experts. We provide an example here to illustrate how this may be done and the identifying power studied. We do not endorse the specific assumptions made here.

Consider the decomposition of COVID-19 episodes into those where the patient does and does not manifest discernible symptoms. Dr. Anthony Fauci, the director of the National Institute of Allergy and Infectious Diseases, has been quoted as saying that the fraction of cases in which the patient is infected but shows no symptoms is "somewhere between 25 and 50 percent." Fauci went on to say "And trust me, that is an estimate. I don't have any scientific data yet."[8]

Supposing it to be correct, Fauci's bound has identifying power when combined with a further assumption. Let $A_d = 1$ or $S_d = 1$ if a person has respectively had an asymptomatic or symptomatic case of COVID-19 by date d. Let each quantity equal zero otherwise. The two categories of illness are mutually exclusive, so $C_d = A_d + S_d$. Hence,

(14)  $P(C_d = 1) = P(S_d = 1) + P(A_d = 1).$

Fauci imposes the assumption

(15)  $P(A_d = 1) = \alpha P(C_d = 1),$ for some $\alpha \in [0.25, 0.5].$

---

[8] See https://www.nytimes.com/2020/04/07/science/coronavirus-uncertainty-scientific-trust.html?action=click&module=Well&pgtype=Homepage§ion=Health

Combining (14) and (15) yields

(16) $\quad P(C_d = 1) = P(S_d = 1)/(1 - \alpha)\;$ for some $\alpha \in [0.25, 0.5]$.

The value of $P(S_d = 1)$ is unknown. However, the existing criteria for testing require the presence of symptoms or close contact with a person known to have been infected. Suppose that persons who are accepted for testing all meet the first criterion. Then the rate of infection with symptoms is greater than or equal to the fraction of the population who are both tested and infected. Assume as earlier that all positive testing results are accurate and that the fraction of inaccurate negative testing results is in the bound $[L_{d10}, U_{d10}]$. Then the fraction who are tested and infected is known to be at least equal to $P(R_d = 1) + L_{d10}P(T_d = 1, R_d = 0)$. Combining this lower bound on $P(S_d = 1)$ with knowledge that $\alpha \geq 0.25$, (16) yields this lower bound on the population infection rate:

(17) $\quad P(C_d = 1) \geq (0.75)^{-1}[P(R_d = 1) + L_{d10}P(T_d = 1, R_d = 0)]$.

This lower bound is $(0.75)^{-1}$ times that in (10), thus improving on it. If one finds bound (17) credible, the final lower bound on $P(C_d = 1)$ is the maximum of the lower bounds in (17) across dates d' ≤ d, as in (13).

2.4. Bounding Rates of Severe Illness Conditional on Infection

Surveillance systems may report several rates of severe illness (V), including hospitalization (H), ICU usage (U), and death (D).[9] The present discussion considers these reports to be accurate. Thus, one may have empirical knowledge of the rates $P(V_d = 1)$ for $V \in \{H, U, D\}$.

---

[9] Hospitalization includes ICU usage.

Surveillance systems do not report rates of severe illness conditional on infection. These have the form

(18)   $P(V_d = 1 | C_d = 1) \;=\; P(V_d = 1, C_d = 1)/P(C_d = 1).$

The numerator $P(V_d = 1, C_d = 1)$ may logically differ from the reported rate $P(S_d = 1)$. This may occur for H and U if some persons hospitalized for COVID-19 are mis-diagnosed. It may occur for D if some reported causes of death are inaccurate. For simplicity, we assume here that such errors do not occur. However, we caution that the assumption may not be realistic.[10]

In the absence of reporting errors, $V_d = 1 \Rightarrow C_d = 1$. Hence,

(19)    $P(V_d = 1 | C_d = 1) \;=\; P(V_d = 1)/P(C_d = 1).$

Given (19), the bound obtained for $P(C_d = 1)$ immediately yields a bound for $P(V_d = 1 | C_d = 1)$. The lower (upper) bound on $P(V_d = 1 | C_d = 1)$ is achieved when $P(C_d = 1)$ takes it upper (lower) bound.

When interpreting rates of severe illness conditional on infection, one should keep in mind that severe cases of COVID-19 may not be apparent as of the date of infection. Many patients begin with mild or no symptoms and develop severe cases a week to two weeks after infection. Hence, the rate of severe illness computed as of a specified date may understate the rate of eventual severe illness.

2.5. Bounding Rates Conditional on Personal Characteristics

The above derivations hold however one defines the population of interest. Application of the bound on the infection rate is possible if one has empirical knowledge of the testing rate and the rate of positive

---

[10] See https://www.washingtonpost.com/investigations/coronavirus-death-toll-americans-are-almost-certainly-dying-of-covid-19-but-being-left-out-of-the-official-count/2020/04/05/71d67982-747e-11ea-87da-77a8136c1a6d_story.html

testing findings for the relevant population. Application of the bound on the rate of severe illness conditional on infection is possible if one additionally has knowledge of the rate of severe illness in the relevant population.

There are both clinical and public health reasons why one would like to know $P(C_d = 1|X)$ and $P(V_d = 1|X, C_d = 1)$ for persons with specified personal characteristics X. For example, it has been thought important to know these rates conditional on the demographic characteristics X = (age, gender, race). Whatever X may be, the bound on $P(C_d = 1|X)$ is this X-specific version of (5):

(20) $\quad P(R_d = 1|X) + L_{dX0}P(T_d = 0|X) + L_{dX10}P(R_d = 0|X, T_d = 1)P(T_d = 1|X) \leq P(C_d = 1|X)$

$$\leq P(R_d = 1|X) + U_{dX0}P(T_d = 0|X) + U_{dX10}P(R_d = 0|X, T_d = 1)P(T_d = 1|X).$$

When computing the bound, one should bring to bear credible X-specific bounds on $P(C_d = 1|X, T_d = 0)$ and $P(C_d = 1|X, T_d = 1, R_d = 0)$. If one imposes a monotonicity restriction conditional on X as in Section 2.2, the bound in (10) is updated similarly as we did in (14) for the bound in (5). The bound on $P(V_d = 1|X, C_d = 1)$ is computable if surveillance additionally reports $P(V_d = 1|X)$.

3. Data

We analyze data from two states in the United States, Illinois and New York, and from Italy. Our data sources are Illinois Department of Public Health (2020), New York State Department of Health (2020), and the Italian Protezione Civile (2020). The data report, among other things, cumulative counts of number of individuals tested and number of positive test results for each day, starting February 24 for Italy, March 10 for Illinois, and March 2 for New York.[11] Our analysis begins from the first day after which all of Illinois, New York, and Italy, had at least one hundred confirmed cases, which is March 16.

For Italy, the data also include the cumulative counts of the number of individuals experiencing severe outcomes (hospitalization, $H_d$; intensive care unit, $U_d$; death, $D_d$). As of the writing of this paper, we do not have access to official data from the Illinois and New York health departments reporting on the time series of these outcomes.

For Illinois and New York, we obtain population size counts using data from Census Bureau (2019) as of July 1, 2019. For Italy, population size is taken from Istituto Nazionale di Statistica (2019) as of January 1, 2019. Our estimates of the quantities in bound (10) are obtained as simple frequency estimators. For example, $P(T_d = 1)$ is computed as the number of tested individuals up to date d divided by population size.

4. Results

Table 1 reports, in columns 2-7, publicly released values of $P(T_d = 1)$ and $P(R_d = 1|T_d = 1)$ for Illinois, New York, and Italy. For several reasons, we caution against comparison of these quantities across states and countries. The criteria for testing and the accuracy of tests may differ across location and may have changed over time. Moreover, the epidemics began at different dates across locations.

The table reveals that from March 16 to April 6, 2020, the fraction of individuals tested increased from 0 to 0.005 in Illinois, from 0.001 to 0.017 in New York, and from 0.002 to 0.012 in Italy. Over the same period, the fraction of positive test results varied (non-monotonically) from 0.092 to 0.195 in Illinois;

---

[11] The documentation for New York indicates that non-positive results are sub-classified into those that are negative and inconclusive. This sub-classification is not made in the documentation for Illinois and Italy. Receipt of test results may take several days or longer. We are not certain how agencies date tests while results are still pending. Our analysis interprets the reported data on number of persons tested to exclude cases where results are pending.

increased monotonically from 0.134 to 0.408 in New York; and varied (non-monotonically) from 0.203 to 0.184 in Italy, with a peak of 0.233.

For Italy, Table 1 also reports, in columns 8-10, the rates of severe outcomes $P(H_d = 1)$, $P(U_d = 1)$, and $P(D_d = 1)$. The data reveal that $P(H_d = 1)$ increased from 0.00021 to 0.00054, and $P(U_d = 1)$ from 0.00003 to 0.00006. The fact that these rates decrease towards the end of the period is due to the reduction in the number of new cases and the increase in the number of recovered individuals that Italy has experienced since April 4. The death rate $P(D_d = 1)$ has increased from 0.00004 to 0.00027.

Table 2 reports the bounds in (13) for Illinois, New York, and Italy, under the monotonicity assumptions presented in Section 2.2. The temporal monotonicity condition is reflected in the fact that, for each state and country, both the lower and upper bounds on $P(C_d = 1)$ weakly increase over time. The widths of the bounds range from 0.455 to 0.516 for Illinois, 0.48 to 0.637 for New York, and remains about 0.51 for Italy throughout the period.

The substantial width of the bounds reflects the fact, previously discussed, that the fraction of tested individuals was very small throughout the period. Nonetheless, the bounds have substantial informational content relative to the extremely wide bounds that would hold if society were to possess no credible information about the infection rate among untested persons. On April 6, 2020, the bounds on the infection rates in Illinois, New York, and Italy respectively are [0.001, 0.517], [0.008, 0.645], and [0.003, 0.510].

We next exemplify how bringing to bear further information from expert opinion, specifically that of Anthony Fauci discussed in Section 2.3, may tighten the bounds. This information does not lower the upper bound on the probability of infection, but it slightly raises the lower bounds by a factor of $(0.75)^{-1}$. Considering again April 6, 2020, the updated bounds are [0.002, 0.517], [0.011, 0.645], and [0.004, 0.510].

Table 3 reports the probability that infected individuals in Italy experience severe health outcomes: $P(H_d = 1|C_d = 1)$, $P(U_d = 1|C_d = 1)$, and $P(D_d = 1|C_d = 1)$, under the monotonicity assumptions for infection rates presented in Section 2.2. Focusing on April 6, 2020, we see that the bound on the probability of being hospitalized if infected is [0.001, 0.172]. The bound on the probability of needing intensive care is narrow, being [0, 0.02]. The fatality rate on April 6 lies in the bound [0.001, 0.086]. It is notable that this upper

bound on fatality is substantially lower than the fatality rate among confirmed infected individuals, which was 0.125 on April 6.

5. Discussion

This paper has used standard methods of partial identification analysis to study two key aspects of the uncertainty that has frustrated attempts to learn the COVID-19 infection rate and rates of severe illness conditional on infection. We have quantified the implications of uncertainty about the infection rate among non-tested persons and about the NPV of the tests in use. The simple analysis of Section 2 shows how available data and maintained assumptions combine to determine the inferences that can logically be drawn. We have used monotonicity assumptions that have strong credibility in the current context. We also have used a conjecture bounding the rate of asymptomatic infection to illustrate how further assumptions having a less firm foundation may be brought to bear, should one find them credible.

We have used data for two American states and for Italy to illustrate application of the analysis. Given that the tested fraction of the population has been very low, one can barely draw any conclusion about the population infection rate without making assumptions that bound the rate of infection in the untested sub-population. Imposing the monotonicity assumptions restricts the population infection rate to bounds that have about width 0.5 in the current covid context.

One naturally may prefer bounds of narrower width. Given the available data, this is logically possible to achieve only if one imposes stronger assumptions with considerable identifying power. We have not reported narrower bounds because we do not immediately see a credible basis to add assumptions that would justify them. Readers who feel that they can motivate stronger assumptions may adapt our analysis to determine their implications.

Among the possibilities for narrowing the bounds, it has often been suggested that we can learn about the prevalence and severity of COVID-19 in one location by observing the experiences of populations in other locations. For example, it has been suggested that the United States can learn from the experience in China, South Korea, and Italy. In these locations the epidemic began earlier and has been handled in different ways. Bringing to bear data from different locations is not helpful per se. It may be helpful if the data are combined with assumptions that enable credible extrapolation across locations. Given such assumptions, the partial-identification sub-literature on *intersection bounds* shows how to proceed formally to tighten inference. See Manski (2020) and Molinari (2020).

To simplify the presentation, we have intentionally abstracted from other potential sources of uncertainty that may further aggravate the inferential problem. We have assumed that persons who recover from the COVID-19 disease become immune and, hence, cannot be infected anew. We have assumed that persons who are tested and receive a negative result are not retested subsequently. We have assumed that hospitals correctly diagnose patients and that public records correctly code causes of death. We caution that these assumptions may not be completely accurate. The partial identification analysis performed in Section 2 may be extended to incorporate these and other further uncertainties.

Departing from conventional practice in applied econometric analysis, we do not refer to the empirical results in Section 4 as "estimates" and we do not provide measures of statistical precision. Instead, we view states and nations as the units of interest rather than as realizations from some sampling process. Measurement of statistical precision requires specification of a sampling process that generates the available data. Yet we are unsure what type of sampling process would be reasonable to assume in this work.

The data we used are exact population counts of tests performed and their results in each location, not observations of samples drawn in the locations. To perform statistical inference, one would have to view the population of each location as the sampling realization of a random process defined on a super-population of alternative population sizes and compositions. See Manski and Pepper (2018) for extended discussion of this matter in a different applied context.

6. Conclusion

      While the bounds we report can be narrowed by imposing stronger assumptions, a more satisfactory way to increase knowledge of the infection rate is to obtain better data. As has been widely recognized, random testing of populations would contribute enormously. Obtaining a firm understanding of the negative predictive value of the tests in use is also important. We urge efforts to progress in both directions.

**Table 1: Probability of being tested, and of receiving a positive test result if tested, and for Italy, probability of severe outcomes**

| Date | Illinois | | New York | | Italy | | | | |
|---|---|---|---|---|---|---|---|---|---|
| | $P(T_d = 1)$ | $P(R_d = 1 \mid T_d = 1)$ | $P(T_d = 1)$ | $P(R_d = 1 \mid T_d = 1)$ | $P(T_d = 1)$ | $P(R_d = 1 \mid T_d = 1)$ | $P(H_d = 1)$ | $P(U_d = 1)$ | $P(D_d = 1)$ |
| 3/16/2020 | 0.000 | 0.092 | 0.001 | 0.134 | 0.002 | 0.203 | 0.00021 | 0.00003 | 0.00004 |
| 3/17/2020 | 0.000 | 0.107 | 0.001 | 0.161 | 0.002 | 0.212 | 0.00025 | 0.00003 | 0.00004 |
| 3/18/2020 | 0.000 | 0.140 | 0.001 | 0.184 | 0.003 | 0.216 | 0.00028 | 0.00004 | 0.00005 |
| 3/19/2020 | 0.000 | 0.134 | 0.002 | 0.218 | 0.003 | 0.225 | 0.00030 | 0.00004 | 0.00006 |
| 3/20/2020 | 0.000 | 0.136 | 0.002 | 0.226 | 0.003 | 0.227 | 0.00031 | 0.00004 | 0.00007 |
| 3/21/2020 | 0.000 | 0.121 | 0.003 | 0.246 | 0.004 | 0.230 | 0.00034 | 0.00005 | 0.00008 |
| 3/22/2020 | 0.001 | 0.125 | 0.004 | 0.266 | 0.004 | 0.229 | 0.00038 | 0.00005 | 0.00009 |
| 3/23/2020 | 0.001 | 0.130 | 0.005 | 0.280 | 0.005 | 0.232 | 0.00040 | 0.00005 | 0.00010 |
| 3/24/2020 | 0.001 | 0.134 | 0.005 | 0.297 | 0.005 | 0.233 | 0.00042 | 0.00006 | 0.00011 |
| 3/25/2020 | 0.001 | 0.131 | 0.006 | 0.305 | 0.005 | 0.229 | 0.00044 | 0.00006 | 0.00012 |
| 3/26/2020 | 0.001 | 0.153 | 0.007 | 0.322 | 0.006 | 0.223 | 0.00047 | 0.00006 | 0.00014 |
| 3/27/2020 | 0.002 | 0.140 | 0.008 | 0.335 | 0.007 | 0.219 | 0.00049 | 0.00006 | 0.00015 |
| 3/28/2020 | 0.002 | 0.137 | 0.009 | 0.345 | 0.007 | 0.215 | 0.00051 | 0.00006 | 0.00017 |
| 3/29/2020 | 0.002 | 0.166 | 0.010 | 0.356 | 0.008 | 0.215 | 0.00052 | 0.00006 | 0.00018 |
| 3/30/2020 | 0.002 | 0.166 | 0.011 | 0.369 | 0.008 | 0.213 | 0.00053 | 0.00007 | 0.00019 |
| 3/31/2020 | 0.003 | 0.170 | 0.011 | 0.379 | 0.008 | 0.209 | 0.00053 | 0.00007 | 0.00021 |
| 4/1/2020 | 0.003 | 0.173 | 0.012 | 0.387 | 0.009 | 0.204 | 0.00054 | 0.00007 | 0.00022 |
| 4/2/2020 | 0.003 | 0.176 | 0.013 | 0.395 | 0.010 | 0.198 | 0.00054 | 0.00007 | 0.00023 |
| 4/3/2020 | 0.004 | 0.185 | 0.015 | 0.401 | 0.010 | 0.193 | 0.00054 | 0.00007 | 0.00024 |
| 4/4/2020 | 0.004 | 0.193 | 0.016 | 0.404 | 0.011 | 0.190 | 0.00055 | 0.00007 | 0.00025 |
| 4/5/2020 | 0.005 | 0.191 | 0.016 | 0.407 | 0.011 | 0.186 | 0.00055 | 0.00007 | 0.00026 |
| 4/6/2020 | 0.005 | 0.195 | 0.017 | 0.408 | 0.012 | 0.184 | 0.00054 | 0.00006 | 0.00027 |

Notes: Population sizes are 12,671,821 for Illinois, 19,453,561 for New York, and 60,359,546 for Italy.

**Table 2: Bounds on infection rate under the testing and temporal monotonicity assumptions**

|  | Illinois | | New York | | Italy | |
| --- | --- | --- | --- | --- | --- | --- |
| Date | LB | UB | LB | UB | LB | UB |
| 3/16/2020 | 0.000 | 0.455 | 0.000 | 0.480 | 0.001 | 0.510 |
| 3/17/2020 | 0.000 | 0.464 | 0.000 | 0.497 | 0.001 | 0.510 |
| 3/18/2020 | 0.000 | 0.472 | 0.000 | 0.511 | 0.001 | 0.510 |
| 3/19/2020 | 0.000 | 0.472 | 0.000 | 0.531 | 0.001 | 0.510 |
| 3/20/2020 | 0.000 | 0.472 | 0.001 | 0.536 | 0.001 | 0.510 |
| 3/21/2020 | 0.000 | 0.472 | 0.001 | 0.547 | 0.001 | 0.510 |
| 3/22/2020 | 0.000 | 0.475 | 0.001 | 0.559 | 0.001 | 0.510 |
| 3/23/2020 | 0.000 | 0.478 | 0.002 | 0.568 | 0.001 | 0.510 |
| 3/24/2020 | 0.000 | 0.479 | 0.002 | 0.578 | 0.002 | 0.510 |
| 3/25/2020 | 0.000 | 0.479 | 0.002 | 0.583 | 0.002 | 0.510 |
| 3/26/2020 | 0.000 | 0.482 | 0.003 | 0.593 | 0.002 | 0.510 |
| 3/27/2020 | 0.000 | 0.482 | 0.003 | 0.601 | 0.002 | 0.510 |
| 3/28/2020 | 0.000 | 0.482 | 0.004 | 0.607 | 0.002 | 0.510 |
| 3/29/2020 | 0.001 | 0.499 | 0.004 | 0.614 | 0.002 | 0.510 |
| 3/30/2020 | 0.001 | 0.500 | 0.005 | 0.622 | 0.002 | 0.510 |
| 3/31/2020 | 0.001 | 0.502 | 0.005 | 0.627 | 0.002 | 0.510 |
| 4/1/2020 | 0.001 | 0.504 | 0.006 | 0.632 | 0.003 | 0.510 |
| 4/2/2020 | 0.001 | 0.506 | 0.006 | 0.637 | 0.003 | 0.510 |
| 4/3/2020 | 0.001 | 0.511 | 0.007 | 0.641 | 0.003 | 0.510 |
| 4/4/2020 | 0.001 | 0.515 | 0.007 | 0.642 | 0.003 | 0.510 |
| 4/5/2020 | 0.001 | 0.515 | 0.008 | 0.644 | 0.003 | 0.510 |
| 4/6/2020 | 0.001 | 0.517 | 0.008 | 0.645 | 0.003 | 0.510 |

24**Table 3: Bounds on the probability of severe illness conditional on infection under the testing and temporal monotonicity assumptions**

| | Italy | | | | | |
|---|---|---|---|---|---|---|
| | Total Hospitalization | | ICU | | Death | |
| Date | LB | UB | LB | UB | LB | UB |
| 3/16/2020 | 0.000 | 0.330 | 0.000 | 0.047 | 0.000 | 0.055 |
| 3/17/2020 | 0.000 | 0.346 | 0.000 | 0.048 | 0.000 | 0.058 |
| 3/18/2020 | 0.001 | 0.341 | 0.000 | 0.046 | 0.000 | 0.061 |
| 3/19/2020 | 0.001 | 0.331 | 0.000 | 0.045 | 0.000 | 0.062 |
| 3/20/2020 | 0.001 | 0.296 | 0.000 | 0.042 | 0.000 | 0.064 |
| 3/21/2020 | 0.001 | 0.287 | 0.000 | 0.040 | 0.000 | 0.067 |
| 3/22/2020 | 0.001 | 0.289 | 0.000 | 0.038 | 0.000 | 0.069 |
| 3/23/2020 | 0.001 | 0.281 | 0.000 | 0.038 | 0.000 | 0.071 |
| 3/24/2020 | 0.001 | 0.275 | 0.000 | 0.037 | 0.000 | 0.074 |
| 3/25/2020 | 0.001 | 0.268 | 0.000 | 0.035 | 0.000 | 0.075 |
| 3/26/2020 | 0.001 | 0.261 | 0.000 | 0.033 | 0.000 | 0.075 |
| 3/27/2020 | 0.001 | 0.254 | 0.000 | 0.032 | 0.000 | 0.078 |
| 3/28/2020 | 0.001 | 0.242 | 0.000 | 0.031 | 0.000 | 0.079 |
| 3/29/2020 | 0.001 | 0.235 | 0.000 | 0.029 | 0.000 | 0.081 |
| 3/30/2020 | 0.001 | 0.228 | 0.000 | 0.029 | 0.000 | 0.083 |
| 3/31/2020 | 0.001 | 0.221 | 0.000 | 0.028 | 0.000 | 0.085 |
| 4/1/2020 | 0.001 | 0.211 | 0.000 | 0.026 | 0.000 | 0.086 |
| 4/2/2020 | 0.001 | 0.201 | 0.000 | 0.025 | 0.000 | 0.086 |
| 4/3/2020 | 0.001 | 0.193 | 0.000 | 0.024 | 0.000 | 0.086 |
| 4/4/2020 | 0.001 | 0.186 | 0.000 | 0.022 | 0.000 | 0.086 |
| 4/5/2020 | 0.001 | 0.178 | 0.000 | 0.021 | 0.001 | 0.086 |
| 4/6/2020 | 0.001 | 0.172 | 0.000 | 0.020 | 0.001 | 0.086 |